\documentclass[12pt]{iopart} 
\usepackage{graphicx} 
 
\begin{document} 
 
\title[]{Troubles for observing the inflaton potential} 
\author{H. P. de Oliveira and C\'{e}sar A.~Terrero-Escalante} 
\address{Departamento de F\'{\i}sica Te\'orica. Instituto de F\'{\i}sica. 
Universidade do Estado do Rio de Janeiro. Rua S\~ao Francisco Xavier 524, 
Maracan\~a, 20559-900 Rio de Janeiro, RJ, Brazil} 
\eads{\mailto{oliveira@dft.if.uerj.br},\ \mailto{cterrero@dft.if.uerj.br}} 
 
\begin{abstract} 
Robustness of the solutions to the inflaton potential 
inverse problem based on the slow-roll approximation is addressed.  
With that aim it is introduced a measure of the difference of the outputs  
obtained using first and second order respectively in the horizon-flow expansion.  
The evolution of this measure is determined by 
a second order linear non-autonomous non-homogeneous differential equation. 
Boundedness of the general solutions to this equation is analyzed. 
It is shown that they diverge for most of the physically meaningful cases. 
Examples for typical inflationary models are presented. 
It is argued that this lack of robustness  
is due to the limitations of the slow-roll expansion 
for probing the scale-dependence of the inflationary spectra. 
\end{abstract} 
 
\begin{flushleft} 
\end{flushleft} 
 
\maketitle 
 
\section{Introduction} 
 
The relatively recent release of the analysis of the first year WMAP data~\cite{Bennett:2003bz}  
and of a significant amount of very precise data  
of the power spectrum of matter distribution at large scales~\cite{Tegmark:2003uf} 
caused a turmoil in the cosmological community. 
The feasibility of obtaining reliable information about our universe  
when it was younger than $10^{-10}$ seconds 
accounts for a large part of that excitement.  
Data confirmed that the primordial power spectrum of 
curvature fluctuations is consistent with a scale-invariant, Gaussian and adiabatic spectrum.  
The simplest and 
most elegant mechanism known to causally produce primordial spectra with 
those properties is cosmological inflation.  
 
In the simplest scenarios with an exit from 
inflation the responsible for the accelerated expansion of the early Universe  
is the {\it inflaton} $\varphi$,  
a single scalar field with equations of motion, 
\begin{eqnarray} 
\label{eq:KGphi} 
\ddot{\varphi} + 3H\dot{\varphi} + V'(\varphi) = 0 \, ,\\ 
\label{eq:Feq} 
H^2 = \frac{1}{3}\left[\frac{\dot{\varphi}^2}{2}+V(\varphi)\right]\, . 
\end{eqnarray} 
Here $V(\varphi)$ is the potential energy density of the inflaton field, 
dot stands for a derivative with respect to cosmic time $t$, 
and prime denotes a derivative with respect to $\varphi$. 
We use Planck units, $8\pi G = c = \hbar = 1$.  
 
The Hubble horizon  
$d_{\rm H} \equiv 1/H$, 
is roughly the size of the region  
where causal processes can take place during one Hubble time $1/H$. 
During inflation the comoving $d_{\rm H}$ decreased 
allowing comoving scales to cross out the causal horizon.  
Physical information 
of the inflaton and of the space-time quantum fluctuations  
contained on those scales 
was effectively `frozen'. 
This way,  
when the scales crossed $d_{\rm H}$ back later during the era of standard (non-accelerated) expansion,  
the whole observable universe was seed  
with statistically uniform curvature perturbations for large-scale structure formation.  
These seeds served also as initial conditions for the evolution of CMBR anisotropies.  
 
The spectra of quantum fluctuations of density and metrics 
during cosmological inflation can be predicted~\cite{Mukhanov:xt}. 
It is required to know the behavior of the Hubble horizon 
as a function of the logarithm of the scale factor, $N\equiv\ln a(t)$. 
In the inflaton scenario it means solving Eqs.(\ref{eq:KGphi}) and (\ref{eq:Feq}). 
For most potentials that is a difficult task that 
can be simplified using the horizon-flow functions which 
are defined recursively 
as~\cite{Schwarz:2001vv}: 
\begin{equation} 
\label{eq:hff} 
\epsilon_{m+1} \equiv {{\rm d} \ln |\epsilon_m|\over {\rm d} N} 
\quad \forall m \geq 0, \quad 
\epsilon_0 = {H(N_{\rm i})\over H(N)}\, , 
\end{equation} 
where $N_{\rm i}$ denotes an arbitrary `initial' moment. The necessary 
condition for inflation to take place is 
$\epsilon_1<1$, and if the weak energy condition and null 
energy condition hold true, $\epsilon_1 \geq 0$.  
In general, for the primordial spectra to be nearly scale invariant, 
it is sufficient to assume $|\epsilon_m|<1$ allowing,  
this way, 
to expand all the interesting inflationary quantities  
in terms of the horizon-flow functions\footnote{ 
Note that according to recent results that condition may be unnecessary~\cite{Makarov:2005uh}.}. 
The \textit{slow-roll approximation} is a particular case where all the $\epsilon_m$ 
are assumed to have the same order of magnitude. 
 
Eqs.(\ref{eq:KGphi}) and (\ref{eq:Feq}) relate the horizon-flow functions 
to the inflaton potential $V(\varphi)$  
and its derivatives with respect to the inflaton field. 
For instance,  
the first two flow functions are given approximately by~\cite{Leach:2002ar}: 
\begin{equation} 
\epsilon_1 \approx 
\frac{1}{2}\left(\frac{V^\prime}{V}\right)^2\, , \qquad 
\epsilon_2 \approx  
2\left[\left(\frac{V^\prime}{V}\right)^2-\frac{V^{\prime\prime}}{V}\right]\,. 
\label{eq:eV} 
\end{equation} 
In turn,  
the behavior of $d_{\rm H}(N)$ near a suitable pivot point $N_*$ 
is described by 
\begin{eqnarray} 
d_{\rm H}(N)&=& d_{\rm H}(N_*)\left[1+\epsilon_1(N-N_*)+ \frac 1 
2(\epsilon_1^2+\epsilon_1\epsilon_2)(N-N_*)^2 
 + \cdots \right]\, . 
\label{eq:dH} 
\end{eqnarray} 
Thus, approximated expressions for the inflationary spectra can be derived. 
To first order for the horizon-flow functions\footnote{ 
Since the order is not the same for all the observables,  
we follow here the convention in Ref.~\cite{Schwarz:2004tz}  
of denoting the order by that of the first term 
in the Taylor parametrization of the primordial spectra.} 
they were found by Stewart and Lyth~\cite{Stewart:1993bc}  
using the slow-roll approximation  
to truncate the argument of Bessel functions  
evaluated at the time of horizon-crossing. 
Their accuracy was tested against numerical solutions in  
Refs.~\cite{Grivell:1996sr,Schwarz:2004tz,Makarov:2005uh}.  
In the same approximation but using Green's functions,  
second order expressions were derived in Refs.~\cite{Gong:2001he,Leach:2002ar} 
and tested by Schwarz and Terrero~\cite{Schwarz:2004tz}.  
The approach based on Bessel functions was then modified~\cite{Schwarz:2001vv,Schwarz:2004tz}  
to derive expressions at any order for two wide classes of inflationary models   
that do not necessarily satisfy the slow-roll condition. 
Up to third order, the corresponding expressions were tested in Ref.~\cite{Schwarz:2004tz}.  
Several other methods were introduced to allow for relaxing the standard slow-roll condition~\cite{Martin:2002vn,Habib:2004kc,Choe:2004zg}.  
In general, their predictions seem to match  
the accuracy of those methods tested against numerical results. 
All these tests yielded that the predictions of inflationary spectra  
using terms up to second order in the horizon-flow expansion  
are accurate enough to match the precision of current and near future observations\footnote{ 
There are a few issues on this regard that will be commented in the last section of this paper.}. 
 
This way, given an inflaton potential as input in Eqs.(\ref{eq:KGphi}) and (\ref{eq:Feq}),  
the horizon-flow functions (\ref{eq:hff}) can be evaluated at any moment,  
the behavior of the Hubble radius can be described by means of (\ref{eq:dH}), 
the spectra of the inflaton and metrics fluctuations accurately predicted, 
used as initial conditions for the calculation  
of the evolution of scalar and tensor modes of anisotropies of the CMBR, 
and the resulting spectra compared with the observational data  
to test the reliability of the given inflationary model.  
 
`Observing' the inflaton potential means solving the related  inverse problem, i.e.,  
that of looking for the functional form of the inflaton potential corresponding to given CMBR spectra. 
Herein, we will refer to this task as the \emph{Inverse Problem for the Inflaton Potential} (IP$^2$). 
Starting with the seminal paper by Hodges and Blumenthal~\cite{Hodges:1990bf}, 
during the last 15 years there were introduced several methods for solving the 
IP$^2$. 
Depending on the way the output is given,  
they can be classified in  
\textit{parametric}~\cite{Hodges:1990bf,Copeland:1993jj,Mangano:1995rh,Mielke:1995pd,Ayon-Beato:2000xx},  
\textit{perturbative}~\cite{Copeland:1993jj,Copeland:1993ie,Lidsey:1995np}, 
\textit{full-numerical}~\cite{Grivell:1999wc}, 
and \textit{stochastic}~\cite{Easther:2002rw} methods. 
 
Assuming that the initial conditions for the calculation of CMBR spectra  
have been already fitted to the observational data, 
what all the methods for solving the IP$^2$ do  
(with the remarkable exception of the full-numerical method)  
is,  
essentially, 
to start by looking for the combinations of $\epsilon_m$ that gives the best fit to those primordial spectra. 
These combinations are constrained by the form and order of the approximated expressions  
for the spectra of density and metrics fluctuations. 
Note that, 
following definitions (\ref{eq:hff}), the number $m$ of horizon-flow functions 
is directly related to the order of the corresponding expressions. 
In its turn, 
the order of the predicted spectra is constrained by the finite precision of data. 
Now, according with expansion (\ref{eq:dH}),  
specifying the set $\{\epsilon_m(N_*)\}$ is equivalent to specifying $d_{\rm H}(N)$.  
So, the unavoidable truncation of $\{\epsilon_m(N_*)\}$ corresponds  
to an incomplete knowledge of the evolution of the Hubble horizon. 
The next step in the IP$^2$ solution is to use Eqs.(\ref{eq:KGphi}) and (\ref{eq:Feq}), 
or the approximations (\ref{eq:eV}), 
to find the functional form of the inflaton potential  
corresponding to that approximated behavior of $d_H(N)$. 
 
The reliability  
of solving the IP$^2$ for a potential  
%matching the available data 
as constrained by measurement errors  
was directly tested in Refs.~\cite{Lidsey:1995np,Copeland:1998fz}. 
There is also a large number of studies  
on the impact of the observational uncertainty 
on the form of the output potential. 
They give us complementary information about possible troubles for observing the potential 
(see Refs.~\cite{Dodelson:1997hr,Leach:2002ar,Schwarz:2004tz,Habib:2004hd,Kadota:2005hv,Kinney:2005in}  
and references therein). 
Though the results are in general encouraging,  
it must be noted that each of them uses in their analysis 
a fixed order for the slow-roll expansion 
of the primordial spectra.   
This is a formal expansion, hence there is not rigorous proof that it converges so, 
there is not certainty about the solution of the inverse problem  
based on this expansion to be robust.  
By \textit{robust} we mean that,  
while increasing the order of the underlying expressions, 
the IP$^2$ solution must converge to a unique functional form.  
%Obviously, t 
This robustness is required in order to make definite statements about 
the high energy physics linked to the observed inflaton potential. 
 
In this sense, the preliminary analysis presented in Ref.~\cite{Terrero-Escalante:2004aw}  
gave us a first cautionary warning about  
the possible existence of troubles for the robust solution of the IP$^2$ 
when the slow-roll approximation is used. 
The aim of this paper is  
to step forward in the analysis of the robustness of the reconstruction of the inflaton potential 
with regards to the convergence of the horizon-flow expansion  
for the spectra of inflationary perturbations under the slow-roll approximation.  
With that in mind, in the next section we will derive the basic equation for our analysis,  
a second order linear non-autonomous non-homogeneous differential equation 
obtained from the comparison between equations for the inflationary perturbations 
to first and second order in the horizon-flow expansion.  
In Sec.\ref{sec:math} we will present mathematical evidence pointing to the 
non-existence of bounded solutions for this equation  
if the time variation of the coefficients is required to be consistent with the inflationary cosmology. 
We reinforce that conclusion in Sec.\ref{sec:phys}, based on the analysis of  
inflationary models belonging to the prototypical classes introduced in Ref.~\cite{Schwarz:2004tz}. 
Finally, in Sec.\ref{sec:disc} we discuss our results. 
 
\section{The basic equation} 
\label{sec:basic} 
 
For finding the functional form of the inflaton potential  
all the methods mentioned in the Introduction   
use the same kind of expressions for the spectra of scalar perturbations. 
Nevertheless, early in the research on the IP$^2$ 
it was realized that the r\^ole of tensor perturbations 
deserves also attention if a unique solution to that problem is desired~\cite{Copeland:1998fz}. 
The information on these modes in the available data is so far 
rather poor~\cite{Bennett:2003bz,Tegmark:2003uf}.  
That is why all the studies  
but the so-called Stewart-Lyth inverse problem~\cite{Ayon-Beato:2000xx} 
avoided to deal with the tensor equations.  
However, the quality of the data on the tensor modes could be radically improved 
if a number of interesting experiments manages to see the primordial light 
(see for instances Refs.~\cite{Ruhl:2004kv}). 
Even more,  
studies based on the parametric method have shown that even a poor information on the spectrum of the 
primordial gravitational waves allows for breaking the degeneracy in the solution of the IP$^2$  
and for obtaining interesting  
results~\cite{Terrero-Escalante:2001ni,Terrero-Escalante:2001du,Terrero-Escalante:2002uj}. 
Particularly, in Ref.~\cite{Terrero-Escalante:2002uj} it was shown how 
using the ratio of the amplitudes of the tensor and scalar perturbations, 
\[ 
r \equiv \alpha \frac{A_T^2}{A_S^2}\, , \qquad \left(\alpha=constant\right)\,, 
\] 
%\begin{equation}  
%\label{eq:r} 
%\end{equation}   
as input for the IP$^2$  
can be a very fruitful way of taking into account the available observational information on both types of spectra 
to observe the convexity of the potential during inflation. 
To do that it is necessary to solve the differential equation  
(recall the definitions (\ref{eq:hff}) for the $\epsilon_m$), 
\begin{equation} 
\label{eq:r_r0} 
\ln\frac{r}{r_0} = \ln\epsilon_1 + C\epsilon_2 + C_1\epsilon_1\epsilon_2 
+ C_2\epsilon_2^2 + C_3 \epsilon_2\epsilon_3 \, ,  
\end{equation} 
where $r_0$ is a constant and 
\begin{eqnarray} 
C\approx -0.7296\, , \\  
C_1 \equiv \left(-\frac{\pi^2}2 + 5 + C \right)\approx -0.6644 \, ,\\ 
C_2 \equiv \left(-\frac{\pi^2}8 + 1\right) \approx -0.2337 \, , \\ 
C_3 \equiv \left(-\frac{\pi^2}{24} + \frac{C^2}2\right) \approx -0.1451\, .	 
\label{eq:Cs} 
\end{eqnarray} 
This equation can be derived from the second order expressions for the amplitudes  
of the scalar and tensor perturbations obtained using the slow-roll approximation~\cite{Gong:2001he,Leach:2002ar}. 
(It can also be derived following the lines described in Ref.~\cite{Terrero-Escalante:2002uj}.) 
 
If the reconstruction of the inflaton potential were robust,  
%with respect to the order of the expressions for the perturbation spectra,  
then we would expect that, 
given $r/r_0$ as input into the IP$^2$, 
the output using up to first order in Eq.~(\ref{eq:r_r0})  
(see Ref.~\cite{Terrero-Escalante:2002uj} for an example of such an output) 
will differ only in small features from the output of the problem with the same input  
but taking into account all terms in (\ref{eq:r_r0}).  
To assess this small departure, 
following Ref.~\cite{Terrero-Escalante:2004aw}, 
we introduce a quantity $\delta(N)$ such that 
\begin{equation} 
^{\rm s}\!\epsilon_1 = \epsilon_1\left(1+\delta\right)\, , 
\label{eq:e1pert} 
\end{equation} 
where the `{\rm s}' denotes the horizon-flow functions in the set to be found if  
second order terms in Eq.~(\ref{eq:r_r0}) were used. 
The standard notation remains for the set found with up to first order terms. 
Next, using definitions (\ref{eq:hff})  
%for the second and third horizon-flow function  
it can be obtained 
\begin{eqnarray} 
\label{eq:e2pert} 
^{\rm s}\!\epsilon_2 &=& \epsilon_2 + \frac{d\delta}{dN}\\ 
\label{eq:e23pert} 
^{\rm s}\!\epsilon_2\,^{\rm s}\!\epsilon_3 &=& \epsilon_2\epsilon_3  
+ \frac{d^2\delta}{dN^2} 
\, ,  
\end{eqnarray} 
 
Now, by definition $\epsilon_1$, $\epsilon_2$ and $\epsilon_3$  
satisfy identically the first order version of Eq.~(\ref{eq:r_r0})  
and we will demand $\delta(N)$ to remain very small as $N$ increases. 
Substituting (\ref{eq:e1pert}), (\ref{eq:e2pert}) and (\ref{eq:e23pert}) into  
(\ref{eq:r_r0}), and keeping only linear terms in $\delta(N)$ and its derivatives, 
the following second order linear non-autonomous non-homogeneous differential equation~\cite{Terrero-Escalante:2004aw}, 
\begin{equation} 
\frac{d^2\delta(N)}{dN^2} + \alpha(N)\frac{d\delta(N)}{dN} - \omega^2\delta(N) = f(N) \, , 
\label{eq:main} 
\end{equation} 
is obtained, where 
\begin{eqnarray} 
\alpha(N) \equiv {{\left(C+C_1\epsilon_1+2C_2\epsilon_2\right)}\over {C_3} } \, ,	\\ 
\label{eq:damping} 
\omega^2 \equiv -{1 \over {C_3} } \approx 6.893 \, , \\ 
\label{eq:stiffness} 
f(N) \equiv -{{\left(C_1\epsilon_1\epsilon_2+C_2\epsilon_2^2+C_3\epsilon_2\epsilon_3\right)}\over {C_3} } \, . 
\label{eq:forcing} 
\end{eqnarray} 
 
For the inverse problem of the inflaton potential to be robust,  
the mildest requirement we could ask for the solution of this equation  
is to be bounded for all possible initial conditions $\delta(N_{\rm i})$. 
As we will show in the next two sections, that seems to be unlikely for all  
$\alpha(N)$ and $f(N)$ physically meaningful. 
 
\section{Mathematical evidence} 
\label{sec:math} 
 
Equations like (\ref{eq:main}) are usually found in mechanical problems like the study of vibrations.  
As in Mechanics,  
$\alpha(N)$ is going to be called here the `damping' of the system,  
$-\omega^2$ the `stiffness',  
while $f(N)$ will be referred as the `forcing'. 
 
First of all, we recall that the formal solution of Eq.~(\ref{eq:main}) is~\cite{Arnold} 
\begin{equation} 
\delta(N)=g^N\delta(N_{\rm i})+\int_{N_{\rm i}}^N \left(g^s\right)^{-1} f(s)ds \, , 
\label{eq:GenSol} 
\end{equation} 
where $g^N:{\rm l\!R}^2\rightarrow {\rm l\!R}^2$ is the linear $(N_{\rm i},N)$-advance mapping for the homogeneous system  
\begin{equation} 
\frac{d^2\delta(N)}{dN^2} + \alpha(N)\frac{d\delta(N)}{dN} - \omega^2\delta(N) = 0 \, . 
\label{eq:mainHom} 
\end{equation} 
As we already mentioned, 
for most of the cosmologically interesting cases, $|\epsilon_m|<1$ for $m>0$.  
According with definitions (\ref{eq:hff}) this means that they vary slowly with $N$.  
So, without lost of generality, we can assume the forcing $f(N)$ to be bounded.  
In that case, boundedness of (\ref{eq:GenSol})  
requires boundedness of solutions to Eq.~(\ref{eq:mainHom}). 
The necessary conditions for the existence of bounded solutions of this equation  
with undetermined coefficients is still an open problem.  
Nevertheless, there are inferences that can be drawn for the case of 
a stiffness strictly negative for all $N$.  
Such a stiffness  
is usually considered as a signal of instability of the trivial solution of the above equation.  
The reason is that for the corresponding to Eq.~(\ref{eq:mainHom}) planar system,  
\[ 
\frac {d{\bf \Delta}}{dN}={\bf \Delta}^T\, 
\begin{array}{|l|l} 0 \quad \hspace{1.cm} 1 \\ \\ \omega^2 \quad -\alpha(N) \end{array}\,\, 
{\bf \Delta} \quad {\rm with}\,\,\, {\bf \Delta}^T \equiv | \delta(N) \,\,\, \frac{d\delta(N)}{dN} |\,, 
%\label{eq:MainSystem}  
\] 
the eigenvalues  
$\lambda^\pm(N) \propto \left( -\alpha(N) \pm \sqrt{\alpha(N)^2+4\omega^2} \right)$ 
point to a saddle-like instability for any given value of $N$.  
If these eigenvalues are slowly changing, then the system is assumed to have no bounded (finite) solutions. A rigorous foundation of this folk theorem is still missing  
together with a precise definition of `slowly changing'. 
However, Lyapunov's direct method reveals that, 
for the existence of bounded solutions to equation (\ref{eq:mainHom}) for all $\delta(N_{\rm i})$,  
the following conditions are close to necessary~\cite{Terrero-Escalante:2005math}: 
the damping $\alpha(N)$ should be positive (at least asymptotically) 
and should increase in an exponential rate. 
As it was already mentioned,  
this last condition is incompatible with all physically meaningful inflationary scenarios. 
 
\section{Physical evidence} 
\label{sec:phys} 
 
To consolidate the analysis in the previous section,  
let us see how the solutions of equation (\ref{eq:main}) behave for physical situations. 
To cover as much as possible the space of inflationary scenarios  
we will use one model from each of the classes introduced in Ref.~\cite{Schwarz:2004tz}.  
We choose that classification because  
it involves only exact expressions.  
Then, the classifying criteria are not dependent on the order of the horizon-flow expansion, 
contrary to what happens in the more frequently used small-field/large-field/hybrid models classification~\cite{Dodelson:1997hr}. 
This was realized first by Schwarz and Terrero~\cite{Schwarz:2004tz} 
and recently confirmed by Kinney and Riotto~\cite{Kinney:2005in}.  
Classification in Ref.~\cite{Schwarz:2004tz} is unambiguous because 
is strictly based on physical criteria, 
namely the behavior of the kinetic and total energy densities of the inflaton field. 
It consists of three classes,  
I) hidden-exit inflation ($\epsilon_2 \leq 0$), 
II) toward-exit inflation with general initial conditions ($0 < \epsilon_2 \leq 2 \epsilon_1$),  
and 
III) toward-exit inflation with special initial conditions ($0 < 2\epsilon_1 < \epsilon_2$). 
 
Monomial potentials like 
\begin{equation} 
V(\varphi) = \lambda \frac{\varphi^n}n 
\label{eq:VII} 
\end{equation} 
with chaotic initial conditions give rise to models of II kind~\cite{Linde:fd}.  
Using relations (\ref{eq:eV}) it is found that for such potentials 
\[ 
\epsilon_1(\varphi) \approx \frac {n^2}2 \frac 1{\varphi^2} \, . 
\] 
To convert from dependence on $\varphi$ into dependence on the efolds number, 
we use  
\begin{equation} 
\frac{d\phi}{dN}=\pm \sqrt{2\epsilon_1} \, , 
\label{eq:dphidN} 
\end{equation} 
what can be derived using definition (\ref{eq:hff}) for $\epsilon_1$ and 
the Friedmann equation (\ref{eq:Feq}) for the inflaton cosmology. 
After doing the conversion it is obtained, 
\[ 
\epsilon_1 \approx \frac n 4 \left(N_f - N\right) \, , \quad \epsilon_2 = \epsilon_3 = \frac 4 n \epsilon_1\, . 
\]  
So, the class criterion is met for $n\geq 2$.  
$N_f$ is the number of efolds by the end of inflation.  
Substituting the above results   
into (\ref{eq:damping}) and (\ref{eq:forcing}) 
the following expressions are obtained, 
\begin{eqnarray} 
\label{eq:dII} 
\alpha(N) = \frac {nC_1 + 8C_2}{4C_3(N_f-N)} + {C \over C3} \, ,\\ 
f(N) = -{{nC_1+4C_2+4C_3} \over {4C_3(N_f-N)^2}} \, .	 
\label{eq:fII} 
\end{eqnarray} 
 
Next, we derive the analogous expressions for class III 
(toward-exit inflation with special initial conditions). 
Typical for this class are models where the inflationary phase starts near a false vacuum. 
An example arising in Superstring Theory is~\cite{Binetruy:ss} 
\[ 
V = V_0 - \frac{m^2\varphi^2}2\,. 
\] 
Here, according to (\ref{eq:eV}), in the slow-roll phase we have,  
\[ 
\epsilon_1(\varphi) \approx \frac 1 2 \,{\frac {{m}^{4}{\phi}^{2}} 
{ \left( {\it Vo}-\frac {m^2\phi^2} 2 \right) ^{2}}} \, , 
\quad 
\epsilon_2 \approx 4 \frac {V_0}{m^2\varphi^2} \epsilon_1 \,. 
\] 
%\begin{equation} 
%\end{equation} 
This kind of models exits inflation gracefully when reaching $\epsilon_1(N_f)=1$ 
so,  
for simplicity,  
we set the condition $\varphi(N_f)=\sqrt{2} {V_0}/{m^2}$  
(implying $V_0\leq 1$) and use (\ref{eq:dphidN}) 
to obtain, 
\[ 
\epsilon_1 \approx \exp{\left[-\frac {2m^2}{V_0}\left(N_f-N\right)\right]} \, ,  
\quad \epsilon_2 = \frac {2m^2}{V_0}\,, 
\quad \epsilon_3 = 0\, . 
\]  
Then, the corresponding damping (\ref{eq:damping}) and forcing (\ref{eq:forcing}) are given by, 
\begin{eqnarray} 
\label{eq:dIII} 
\alpha(N) = \frac {C_1}{C_3}\exp{\left[-\frac {2m^2}{V_0}\left(N_f-N\right)\right]}  
+ \frac{4C_2}{C_3} \frac {2m^2}{V_0} + {C \over C3} \, ,\\ 
f(N) = -\frac {2C_1}{C_3}\frac {m^2}{V_0}\exp{\left[-\frac {2m^2}{V_0}\left(N_f-N\right)\right]}  
- \frac{4C_2}{C_3} \frac {m^4}{V_0^2} \, .	 
\label{eq:fIII} 
\end{eqnarray} 
 
Finally, typical examples from class I  
(hidden-exit inflation) 
are hybrid models  
where cosmic acceleration is driven by the inflaton 
and, in order to end inflation,  
there is yet another field which finally drives the vacuum energy $V_0$ to zero. 
Our example is the hybrid model with effective potential~\cite{Linde:1991km} 
\[ 
V = V_0 + \frac{m^2\varphi^2}2 \,. 
\] 
In this case from (\ref{eq:eV}) we obtain 
\[ 
\epsilon_1(\varphi) \approx \frac 1 2 \,{\frac {{m}^{4}{\phi}^{2}} 
{ \left( {\it V_0}+\frac {m^2\phi^2} 2 \right) ^{2}}} \, , 
\quad 
\epsilon_2 \approx - 4 \frac{V_0}{m^2\varphi^2} \epsilon_1 \,. 
\] 
%\begin{equation} 
%\end{equation} 
Similarly to the previous example,  
but now in the opposite regime, 
to simplify the calculations we assume the initial conditions 
for the inflationary era\footnote{ 
There are uncountable others choices that will led to similar results 
and we should expect our analysis to be independent on the initial conditions for inflation.} 
$\epsilon_1(N_{\rm i})=1$, $\varphi(N_{\rm i})=\sqrt{2} {V_0}/{m^2}$, 
what after using (\ref{eq:dphidN}) leads to 
\[ 
\epsilon_1 \approx \exp{\left[-\frac {2m^2}{V_0}\left(N-N_{\rm i}\right)\right]} \, ,  
\quad \epsilon_2 = -\frac {2m^2}{V_0}\,, 
\quad \epsilon_3 = 0\, . 
\]  
The corresponding damping and forcing are then given by, 
\begin{eqnarray} 
\label{eq:dI} 
\alpha(N) = \frac {C_1}{C_3}\exp{\left[-\frac {2m^2}{V_0}\left(N-N_i\right)\right]}  
- \frac{4C_2}{C_3} \frac {2m^2}{V_0} + {C \over C3} \, ,\\ 
f(N) = \frac {2C_1}{C_3}\frac {m^2}{V_0}\exp{\left[-\frac {2m^2}{V_0}\left(N-N_i\right)\right]}  
- \frac{4C_2}{C_3} \frac {m^4}{V_0^2} \, .	 
\label{eq:fI} 
\end{eqnarray} 
 
\begin{figure} 
\centerline{\includegraphics[width=0.55\linewidth]{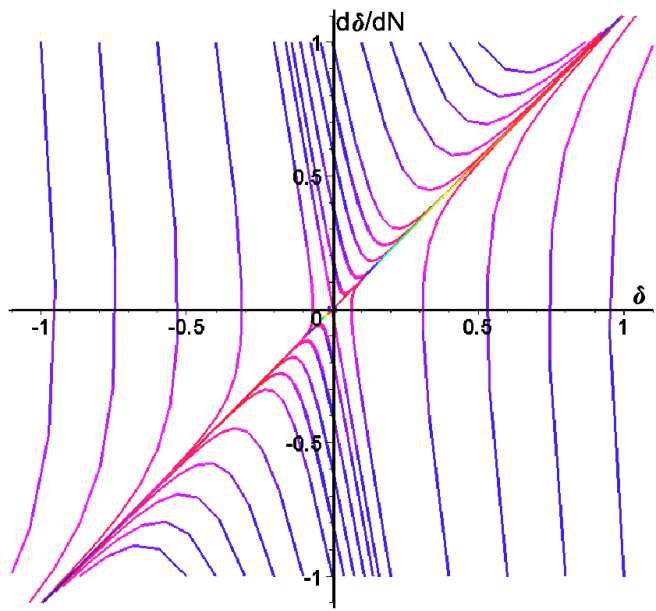}} 
\caption{Plot of the solutions of equation (\ref{eq:main}) for the chaotic model  
$V(\varphi) = \lambda \varphi^4/4$.  
Each curve corresponds to a different solution starting at $N_i=0$ (blue) 
and running towards $N_f=100$ (red).}  
\label{fig:fig1} 
\end{figure}

After substituting expressions (\ref{eq:dII}) and (\ref{eq:fII}),  
and putting the numbers (\ref{eq:Cs}), 
one can look for numerical solutions to Eq.~(\ref{eq:main}). 
In Fig.\ref{fig:fig1} is shown the phase-space   
for a potential given by Eq.~(\ref{eq:VII}) with $n=4$.  
%%%%%%%%%%%%%%%%%%%%%%%%Figure 1%%%%%%%%%%%%%%%%%%%%%%%%%%%%%%% 
%%%%%%%%%%%%%%%%%%%%%%%%%%%%%%%%%%%%%%%%%%%%%%%%%%%%%%%%%%%%%%%%%%%%%%%%%%%% 
Note that the value of $n$ does not modify the qualitative behavior of these solutions, 
while the value of $\lambda$ does not play any r\^ole at all.  
Hence, we can conclude that for this kind of models  
the solution obtained to second order in the horizon-flow functions 
will typically depart from the lowest orders solutions. 
 
The phase diagrams for the damping and forcing given correspondingly by 
(\ref{eq:dIII}) and (\ref{eq:fIII}), or by (\ref{eq:dI}) and (\ref{eq:fI}), 
are topologically equivalent to that in figure \ref{fig:fig1}. 
Therefore, it is confirmed the result in the previous section 
about solutions to Eq.~(\ref{eq:main}) diverging around a saddle-point near the origin,  
as the generic dynamical pattern for those cases which are interesting from the point of view 
of inflationary cosmology.

\section{Discussion} 
\label{sec:disc} 
 
The rigorous conclusion to be drawn from the results presented in this paper is that 
solutions of the inverse problem for the inflaton potential  
obtained using expressions up to second order in the slow-roll approximation 
will be generically very different  
from those solutions obtained using lower order expressions. 
 
It may happen that including higher orders helps to get rid of this trouble.  
However, the order of the expressions used to predict the inflationary spectra 
is determined by the accuracy of the data  
from the observations of the CMBR anisotropies and large-scale matter distribution.  
As it has been pointed out, 
second order expressions will indeed match the precision of current and near future measurements~\cite{Schwarz:2004tz,Makarov:2005uh}. 
In order words,  
in the expressions for the amplitudes of the inflationary spectra,  
horizon-flow functions with orders higher than $2$ 
would not contain any useful information. 
Therefore, including such orders would be irrelevant for the output of the inverse problem.  
 
Note now that equation (\ref{eq:main}) has trivial solution $\delta(N)=0$ only for the 
cases of de Sitter (where $\epsilon_m=0$ for $m>0$ ) 
and power-law inflation~\cite{Lucchin:1984yf}. 
In this last model $a\propto t^p$ (with a constant $p\gg 1$), 
implying $d_{\rm H}\propto\exp(N/p)$, $\epsilon_1=1/p$, $\epsilon_m=0$ for $m>1$ 
and $V(\varphi)=\Lambda\exp(B\varphi)$.  
Here $B$ is a real constant which is zero for a universe dominated by a cosmological constant. 
Therefore, $\delta(N)$ actually measures the depart from an exponential 
of the functional form of the output potential.  
Power-law inflation is the unique model which predicts exact power laws  
for the amplitudes of both the tensor and scalar spectra~\cite{Abbott:1984fp}, i.e., 
$A(k)=A_*k^n$, where $k$ is the comoving wavenumber and $n$ is a constant called the spectral index.   
Moreover, during power-law inflation the scalar index $n_{\rm S}$ is equal to the tensor $n_{\rm T}$,  
what implies a constant ratio of tensor to scalar amplitudes, $r$. 
Conversely, in Ref.~\cite{Terrero-Escalante:2001ni} it was proved that 
using power-law spectra as input in the inverse problem renders the exponential potential as the unique output. 
From this point of view, $\delta(N)$ measures the depart  
of the behavior of the inflationary spectra from a power-law. 
We can conclude that 
the essence of the troubles for observing the inflaton potential 
is the weak signal in the CMBR of scale-dependence of the primordial spectra  
and the difficulties to extract that information from data  
using expressions for the inflationary perturbations based on the slow-roll approximation. 
 
As a matter of fact, 
we were able to clarify here a point which evidences can be traced back in the existing research literature. 
In 1995, Lidsey et al.~\cite{Lidsey:1995np} showed that the results for the spectra obtained 
to first order in the slow-roll approximation 
by Stewart and Lyth~\cite{Stewart:1993bc}  
could be understood as an expansion about the exact solution for power-law inflation. 
Two years later, Wang et al.~\cite{Wang:1997cw} analyzed the implications that finite orders 
in such an expansion have for precisely predicting the inflationary spectra. 
Limitations pointed out were linked to the Bessel and horizon-crossing approximations. 
Martin et al.~\cite{Martin:1999wa} quantified these limitations  
by comparing a power-law parametrization for the amplitudes with a more general Taylor series parametrization, 
and by analyzing the impact of the variation of the pivot scale on the accuracy of predictions.  
The larger the scale-dependence of the spectra, the bigger is this impact  
(see also Ref.~\cite{Kinney:2005in}).  
%Their results were confirmed recently by Kinney and Riotto \cite{Kinney:2005in}. 
 
The combination of the Bessel and slow-roll approximations  
is no longer needed to calculate the primordial spectra,  
and the power-law parametrization has been extended to included the running of the scalar index. 
However, the slow-roll condition remains on the basis of the expressions that usually link 
the Taylor parametrized spectra and the inflationary predictions. 
There is evidence that the above mentioned limitations still persist. 
Comparing approximated and numerical predictions for three given models, 
in Ref.~\cite{Schwarz:2004tz}  
it was noted that  
increasing the order of the horizon-flow functions beyond the quadratic  
in every included term of the Taylor parametrization of primordial spectra 
does not significantly improve the accuracy of the approximations. 
In fact, adding terms in that parametrization 
actually seems to be most important for increasing the precision of the predictions. 
However, the number of terms in the parametrization  
depends on the capability of the observations to measure  
a significant departure from scale-invariance in the primordial spectra. 
As a final evidence, we recall that  
using first order expressions for $r$ and $n_{\rm S}$, 
Dodelson et al. were able to discriminate among inflationary models with 
decreasing and increasing kinetic energy density 
by distributing them in the $(r,n_{\rm S})$ plane~\cite{Dodelson:1997hr}. 
As it was noted in Ref.~\cite{Schwarz:2004tz}, 
including higher order corrections makes impossible to distinguish between these models   
on the basis of a $r-n_{\rm S}$ plot, unless the running of the spectral 
index is known or constrained to be very small. 
The same conclusion was recently made by Kinney and Riotto~\cite{Kinney:2005in}. 
 
Limitations from theoretical and observational sides  
to account for the scale-dependence of the primordial spectra 
have been misunderstood as 
generic predictions of inflation. 
That this scale-dependence is in the edge of  
observational capabilities does not means  
that all the successful inflaton potentials must resemblance an exponential 
during the inflationary era,  
but just that we are unable so far to observe the difference. 
The findings by Hoffman and Turner~\cite{Hoffman:2000ue} 
about power-law inflation as an $\alpha$-attractor  
(attractor in the future) 
of the flow in the space of inflationary models 
were constrained by the order of the expressions in their analysis.  
As it was shown in Ref.~\cite{Terrero-Escalante:2002sd},  
departure from power-law behavior is nullified 
whenever the leading order is used. 
To first order, for instance, it remains true  
that the inflationary dynamics yielding a constant tensor index $n_{\rm T}$ 
has power-law inflation as an $\alpha$-attractor~\cite{Terrero-Escalante:2001ni}.  
Nevertheless, 
it is also true that the inflationary dynamics characterized by a constant ratio  
%of scalar and tensor perturbations,  
$r$, 
has power-law inflation as an $\omega$-attractor  
(attractor in the past or repellor)~\cite{Terrero-Escalante:2001du}. 
Furthermore, in Ref.~\cite{Terrero-Escalante:2002uj} it was shown how a minimal scale-dependence for $r$ 
(and information on such a scale-dependence can be drawn from the difference between both spectral indices) 
allows for models with power-law inflation as a transient regime. 
Based on these facts,  
we conjecture here that,  
if using higher orders in the slow-roll expansion 
for the analysis of the inflationary flow, 
power-law inflation still arises as a fixed point,  
then it will do it like a saddle-point,  
consistently with the results presented in this paper.     
 
Summarizing, either the slow-roll expansion does not converge or it does so slowly 
that a large number of terms in this expansion is required to probe the scale dependence of the primordial spectra. 
Nevertheless, information on this dependence seems to be a necessary ingredient for a robust solution to the  
inflaton potential inverse problem. This conclusion should encourage  
further development of the full-numerical method  
for the resolution of this inverse problem~\cite{Grivell:1999wc}, 
as well as of other analytical methods,  
for instance, 
those based on the WKB~\cite{Martin:2002vn} approximation,  
on the uniform approximation~\cite{Habib:2004kc}, 
and on the inverse-scattering theory~\cite{Habib:2004hd}.

\section*{Acknowledgements} 
 
C.~A.~T-E. thanks the Department of Mathematics and Statistics 
of the Dalhousie University for great hospitality, 
and Alan Coley for very useful suggestions. 
C.~A.~T-E. research was supported by grant CNPq/CLAF-150548/2004-4. 
H.~P.~de Oliveira acknowledges CNPq financial support.

\end{document}